\documentclass[aps,prb,twocolumn,amssymb,superscriptaddress,showpacs]{revtex4-1}
\usepackage{float}
\usepackage{xcolor}
\usepackage{graphicx}
\usepackage{dcolumn}
\usepackage{bm}
\usepackage{hyperref}
\usepackage{color}
\usepackage{amsmath}
\usepackage{amssymb}
\usepackage{amsfonts}
\usepackage{setspace}

\begin{document}

\title{Quantifying proximity-induced superconductivity from first-principles calculations}

\author{Yunhao Li}
\affiliation{Institute for Advanced Study, Tsinghua University,
Beijing 100084, China}

\author{Zimeng Zeng}
\affiliation{Institute for Advanced Study, Tsinghua University,
Beijing 100084, China}
\affiliation{State Key Laboratory of Low Dimensional Quantum
Physics, Department of Physics, Tsinghua University, Beijing
100084, China}

\author{Jizheng Wu}
\affiliation{School of Materials Science and Engineering, Beihang University, Beijing 100191,  China}

\author{Chen Si}
\email{E-mail: sichen@buaa.edu.cn}
\affiliation{School of Materials Science and Engineering, Beihang University, Beijing 100191,  China}

\author{Zheng Liu}
\email{E-mail: zheng-liu@tsinghua.edu.cn}
\affiliation{Institute for Advanced Study, Tsinghua University,
Beijing 100084, China}
\affiliation{State Key Laboratory of Low Dimensional Quantum
Physics, Department of Physics, Tsinghua University, Beijing
100084, China}

\begin{abstract}
Proximity induced superconductivity with a clean interface has attracted much attention in recent years. We discuss how the commonly-employed electron tunneling approximation can be hybridized with first-principles calculation to achieve a quantitative characterization starting from the microscopic atomic structure. By using the graphene-Zn heterostructure as an example, we compare this approximated treatment to the full \textit{ab inito} anisotropic Eliashberg formalism. Based on the calculation results, we discuss how superconductivity is affected by the interfacial environment.   
\end{abstract}

\maketitle

\setstretch{1}

\section{Introduction}

Since the establishment of the BCS theory\cite{10,11}, the occurrence of superconductivity (SC) in non-superconducting materials (N) placed in proximity to a superconductor (S) has been studied by various semiempirical approximations\cite{5,6,12}. Historically, the dirty interface was better modelled, for which the detailed interfacial structure is less important and the motion of the superconducting electrons can be described by a simplified diffusion equation\cite{6,13}. In contrast, for the clean NS heterostructure, an atomic characterization of the interfacial coupling is more complicated. 

In the past two decades, the first-principles calculation within the framework of density functional theory (DFT)\cite{14,15} has reached a status to reliably describe not only the normal states of a wide range of materials\cite{16,17}, but also conventional superconductivity mediated by phonons\cite{20,21}. Calculating electron-phonon couplings (EPCs) from first principles is rapidly reaching maturity, thanks to the development of density functional perturbation theory (DFPT)\cite{18,19}. In addition, the computational cost of evaluating EPCs on a dense mesh of the Brillouin zone is significantly reduced by the efficient first-principles interpolation technique based on maximally localized Wannier functions (MLWF)\cite{21}. This progress makes quantitative and predictive calculations on the interface superconductivity possible.

Meanwhile, the development of microfabrication technologies renders manufacturing N-S interfaces in a controllable way at nanoscales\cite{8,9}, and discussions on the clean-limit proximity effect recevied revived interests, especially regarding N-layers containing special microscopic electron structures, such as relativistic band dispersion, nontrivial band topology and magnetism\cite{1,2,3,4,5,6,7}. 

This Article aims to demonstrate a general strategy to quantify proximity-induced superconductivity from first principles. We use the graphene-Zn  heterostructure as an example to calculate and compare the performance of different treatments of the proximity effects, such as the electron tunneling approximation and complete DFT+DFPT calculations.

\section{Theoretical framework}

Let us consider three atomic structures for first-principles calculations: (i) a N slab; (ii) a S slab; and (iii) a NS heterostructure by combining the two. For practical purposes, we expect that the two slabs have commensurate or nearly commensurate surfaces, so the heterostructure can be constructed in a computationally feasible supercell.

Applying DFT and DFPT calculations to these structures renders the following description:
\begin{eqnarray}\label{eq:Halpha}
H^{\alpha}=H^{\alpha}_e+H^{\alpha}_{ph}+H^{\alpha}_{e-ph},
\end{eqnarray}
with $\alpha$=N, S and NS. The electron and phonon Hamiltonians ($H_{e}$ and $H_{ph}$) are readily diagonalized by DFT and DFPT, and the EPC Hamiltonian ($H_{e-ph}$) is parameterized in the momentum space:
\begin{eqnarray}
\begin{cases}
&H^{\alpha}_e=\sum_{n\mathbf{k}}{\varepsilon_{n\mathbf{k}}^\alpha c_{\alpha n\mathbf{k}}^\dagger}c_{\alpha n\mathbf{k}}\\
&H^{\alpha}_{ph}=\sum_{\upsilon ,q}{\hbar \omega^\alpha_{\nu\mathbf{q}} b_{\alpha\upsilon,\mathbf{q}}^{\dagger}b_{\alpha\upsilon ,\mathbf{q}}} \\
&H^{\alpha}_{e-ph}=\sum_{n,m,\mathbf{k,q},\upsilon}g_{n\mathbf{k},m\mathbf{k+q}}^{\alpha\upsilon}c_{\alpha m\mathbf{k+q}}^{\dagger}c_{\alpha n\mathbf{k}}\\
&\times(b_{\alpha \upsilon,\mathbf{q}}^{\dagger}+b_{\alpha \upsilon ,-\mathbf{q}}), \\
\end{cases}
\end{eqnarray}
in which $\varepsilon_{n\mathbf{k}}^\alpha$ ($\omega^\alpha_{\nu\mathbf{q}}$) and $c_{\alpha n\mathbf{k}}$ ($b_{\alpha\upsilon,q}$) denote the electronic (phonon) eigenenergies and annihilation operators acting on the eigenstates indexed by the in-plane lattice momentum $\mathbf{k}$ ($\mathbf{q}$) and an additional band label $n$ ($\upsilon$). $g_{n\mathbf{k},m\mathbf{k+q}}^{\alpha\upsilon}$ is the EPC coefficient. It is understood that when magnetism and spin-orbit coupling are taken into account, $n$ should also index the spin degree of freedom. 

All the proximity effects (PEs) are in principle encoded in:
\begin{eqnarray}\label{eq:HPE}
H^{PE} \equiv H^{NS}-H^N-H^S,
\end{eqnarray}
which can be divided according to Eq. (\ref{eq:Halpha}) into:
\begin{eqnarray}\label{eq:Hsep}
H^{PE}= H_e^{PE}+H_{ph}^{PE}+H_{e-ph}^{PE}.
\end{eqnarray}
Among the three terms, $H_e^{PE}$ is in many cases treated as the main driver\cite{4,5,6}, which not only helps simplify the calculation, but also provides an pedagogical understanding on the superconductivity induced in N via electron tunneling. We will first discuss how to hybridize first-principles calculation with this electron tunneling approximation, and then switch to a full \textit{ab initio} description.

We note that a prerequisite for the discussions below is that the DFT and DFPT descriptions are adequate for the consisting slabs. The only effect of the residual electron-electron interaction is presumed to be an isotropic reduction of the phonon-mediated e-e attraction, as parameterized by a single dimensionless Coloumb pseudopotential $\mu^*$. We do not consider cases violating the Migdal theorem either.

\subsection{Electron tunneling approximation}
Many model studies on the proximity effect directly start from coupling a pure electron Hamiltonian for the N layer to a Bogoliubov-de Gennes Hamiltonian for the S layer. Referring back to the first-principles formalism, this treatment can be rephrased as:

(i) Select electron tunneling as the dominant PE:
\begin{eqnarray}\label{eq:HPE1}
H^{PE} \approx H_e^{PE}.
\end{eqnarray}

(ii) For $H^N$, the phonon effect is assumed to be negligible:
\begin{eqnarray}\label{eq:HN1}
H^N\approx H^N_e.
\end{eqnarray}

(iii) $H^S$ is simplified into a single-band BCS Hamiltonian:
\begin{eqnarray}\label{eq:HS1}
H^S&\approx& \sum_{\mathbf{k}\sigma}{\varepsilon_{\mathbf{k}\sigma}^S c_{S\mathbf{k}\sigma}^\dagger}c_{S\mathbf{k}\sigma} \\ \nonumber
&-&V^S\sum_{\mathbf{k}\mathbf{k'}} c_{S\mathbf{k},\uparrow}^\dag c_{S\mathbf{-k},\downarrow}^\dag c_{S\mathbf{-k'},\downarrow} c_{S\mathbf{k'},\uparrow},
\end{eqnarray}
in which $V^S$ is the averaged pairing potential on Fermi surface (FS). $V^S$ modified by the FS density of states (DOS) ($N_F$) is associated to the dimensionless EPC strength ($\lambda$) routinely computed from DFPT:
\begin{eqnarray}\label{eq:VS1}
V^S =\frac{\lambda^S}{N^S_F}=\langle\sum_\nu\frac{2|g_{\mathbf{k},\mathbf{k+q}}^{S\upsilon}|^2}{\omega^S_{\nu\mathbf{q}}}\rangle_{FS}.
\end{eqnarray}

We first consider the effect of $H_e^{PE}$ [Eq.(5)]. The electronic eigenstates of the heterostructure ($|NS, n\mathbf{k}\rangle$) form a new basis by hybridizing $|N, n\mathbf{k}\rangle$ and $|S, \mathbf{k}\rangle$. Then, rotating the EPCs and the effective pairing potential to this new basis gives:
\begin{eqnarray}\label{eq:gNS1}
g^{NS,\nu_S}_{n\mathbf{k},\mathbf{mk+q}}&\approx& \\ \nonumber 
&&\langle NS, n\mathbf{k}| S, \mathbf{k}\rangle g^{S, \nu_S}_{\mathbf{k}, \mathbf{k+q}} \langle S, \mathbf{k+q}| NS, m\mathbf{k+q}\rangle
\end{eqnarray}
and
\begin{eqnarray}\label{eq:VNS1}
V^{NS}_{n\mathbf{k},m\mathbf{k'}}&\approx& V^S |\langle NS, n\mathbf{k}| S, \mathbf{k}\rangle|^2 |\langle S, \mathbf{k'}| NS, m\mathbf{k'}\rangle|^2\\ \nonumber 
&=&\frac{\lambda_S}{N_F^S}w^S_{n\mathbf{k}}w^S_{m\mathbf{k'}}
\end{eqnarray}
The projection weight appearing in the last line is defined by: $w^S_{n\mathbf{k}}\equiv|\langle NS, n\mathbf{k}| S, \mathbf{k}\rangle|^2$. Since we have assumed the commensurate condition, all the momenta refer to a common super Brillouin zone. There is no overlap between states with different $\textbf{k}$s.

For a multiband SC, we extend this formula by approximating  $w^S_{n\mathbf{k}}$ as the total weight of $|NS,n\mathbf{k}\rangle$ projected into the S slab. We can also include the EPC contributions from the N slab, writing:
\begin{eqnarray}
V^{NS}_{n\mathbf{k},m\mathbf{k'}}&\approx&\frac{\lambda_S}{N_F^S}w^S_{n\mathbf{k}}w^S_{m\mathbf{k'}}+\frac{\lambda_N}{N_F^N}w^N_{n\mathbf{k}}w^N_{m\mathbf{k'}},
\end{eqnarray}
in which $\lambda^N$ is the dimensionless EPC strength defined on the FS of the N-slab.

The approximated $V^{NS}_{n\mathbf{k},m\mathbf{k'}}$ can thus be organized into a 2$\times$2 block matrix according to the projection weights:
\begin{eqnarray}\label{VNS}
V^{NS}=
\left(
\begin{array}{cc}
    V^{NS}_{N\rightarrow N} & V^{NS}_{N\rightarrow S}  \\
    V^{NS}_{S\rightarrow N} & V^{NS}_{S\rightarrow S}
\end{array}
\right).
\end{eqnarray}

When the coefficients of $V^{NS}$ do not vary drastically within each block, it is plausible to perform block average. In analogy to the method used for a two-gap superconductor, e.g. MgB$_2$\cite{22,35,36,37}, a 2$\times$2 dimensionless EPC matrix can be defined:
\begin{eqnarray}\label{eq:Lambda}
\Lambda=
\left(
\begin{array}{cc}
    V^{NS}_{N\rightarrow N}N_F^N & V^{NS}_{N\rightarrow S}N_F^N  \\
    V^{NS}_{S\rightarrow N}N_F^S & V^{NS}_{S\rightarrow S}N_F^S
\end{array}
\right),
\end{eqnarray}
in which $N_F^\alpha$ is the contribution to the FS DOS from the corresponding block. The largest eigenvalue of $\Lambda$, denoted as $\lambda_{max}$, can then be plugged into the semiempricial McMillan-Allen-Dynes formula \cite{23} to predict $T_c$ of this hybrid system:
\begin{eqnarray}\label{eq:MAD}
k_BT_c=\frac{\hbar\omega_{log}}{1.2}\exp[-\frac{1.04(1+\lambda_{max})}{\lambda_{max}-\mu^*(1+0.62\lambda_{max})}],
\end{eqnarray}
where $\omega_{log}$ is a logarithmic average of the phonon frequencies. The ratio between the superconducting gaps in the N (proximitized gap) and S (intrinsic gap) layers can be estimated by the eigenvector corresponding to $\lambda_{max}$. We will elaborate on these details in Sec. III based on a concrete example. 

The great advantage of the electron tunneling approximation is that Eqs.(\ref{eq:HPE1}) refers to the DFT data only, while the expensive EPC calculation is restricted to isolated S and N slabs, which significently reduces the computational complexity.

\subsection{Full \textit{ab initio} treatment}

If a complete DFT+DFPT calculation on the heterojunction is attainable, $V^{NS}_{n\mathbf{k},m\mathbf{k'}}$  can be obtained without approximation. Plus, the semi-empirical McMillan-Allen-Dynes formula can be replaced by the anisotropic and frequency-dependent Migdal-Eliashberg equations\cite{20,21}:
\begin{eqnarray}\label{eq:el}
\begin{cases}
	&Z({\bf k},i\omega_n) = 1 + \frac{\pi T}{N_{\rm F}\omega_n} \sum_{{\bf k}'n'}\frac{\omega_n'}{\sqrt{\omega_n'^2 + \Delta^2({\bf k}',i\omega_n')} } \\
	&\times\lambda({\bf k},{\bf k}',n\!-\!n') \delta(\epsilon_{{\bf k}'})\\
	&Z({\bf k},i\omega_n) \Delta({\bf k},i\omega_n)=\frac{\pi T}{N_{\rm F}} \sum_{{\bf k}' n'}\frac{ \Delta({\bf k}',i\omega_n') }{ \sqrt{\omega_n'^2+\Delta^2({\bf k}',i\omega_n')} }\\
	&\times\left[ \lambda({\bf k},{\bf k}',\!n-\!n')-N_{\rm F}V({\bf k}-{\bf k}')\right] \delta(\epsilon_{{\bf k}'}),\\
\end{cases}
\end{eqnarray}
with:
\begin{eqnarray}\label{eq:epcel}
	\begin{cases}
		\lambda({\bf k},{\bf k}',n - n') = \int_{0}^{\infty} d\omega
		\frac{2\omega}{(\omega_n - \omega_n')^2+\omega^2}\alpha^2F({\bf k},{\bf k}',\omega)\\
		\alpha^2F({\bf k},{\bf k}',\omega) = N_{\rm F} \sum_{\nu} |g_{\mathbf{k},\mathbf{k+q}}^{\upsilon}|^2 \delta(\omega-\omega_{{\bf k}-{\bf k}',\nu}),\\
	\end{cases}
\end{eqnarray}
in which $\Delta$,$Z$ and $\omega_n$ represent the superconducting gap, renormalization function and fermion Matsubara frequencies. Solving Eq. (\ref{eq:el}) self-consistently for the heterostructure reduces approximations to the least level within the DFT and DFPT formalism.

It is worth mentioning that besides $H_e^{PE}$, $H_{ph}^{PE}$ and $H_{e-ph}^{PE}$ may also play an important role, e.g. via phonon renormalization and interfacial phonon scattering, which is captured by the full \textit{ab initio} treatment. The relative importance of these different mechanisms could be strongly system dependent, which is hard to decide \textit{a priori} without microscopic calculations. Whenever possible, a crosscheck between the electron tunneling approximation and the full \textit{ab initio} treatment will be helpful for understanding the origin(s) of the proximity-induced superconductivity.

\section{Case study: graphene on Zn}

As an example to apply the general framework, we consider the graphene-superconductor heterojunction, which has led to useful applications, such as photon detectors\cite{24} and Cooper pair splitters\cite{25}, and motivates a variety of theoretical proposals to achieve exotic superconducting phases\cite{26}. We note that while experiments usually apply an external voltage and measure the supercurrent injected in graphene, here, we focus on the equilibrium SC state in the heterojunction.

\subsection{Numerical setup}

A 6-layer Zn (001) slab is chosen as the superconducting substrate. The experimental critical temperature of Zn is reported to be T$_c$=0.79 K\cite{27}.
We choose Zn mainly for a good lattice match to graphene. Fixing the in-plane lattice constants of the computational supercell according to the fully relaxed Zn bulk's parameters $a$=$b$=4.97 Bohr (cf. the experimental value $a$=$b$=5.04 Bohr\cite{29}) introduces about 7$\%$ tensile strain to the graphene. According to our previous works on graphene\cite{33}\cite{34}, tensile strain tends to soften the phonons and enhance the EPC strength, but 7$\%$ is not sufficient to induce intrinsic superconductivity within a reasonable carrier density range. The atomic structure of the grephene-Zn heterojunction is shown in Fig. 1, including a 28 $\AA$ thick vaccuum layer normal to the 2D surface. The Zn slab is cleaved from a fully relaxed bulk structure, and the Zn-C interfacial spacing is determined by minimizing the total energy. We do not consider surface corrugation or additional structural reconstruction, so a minimal unit cell containing two C atoms and six Zn atoms can be used with periodic boundary conditions.

\begin{figure}
	\includegraphics[width=1.0\columnwidth]{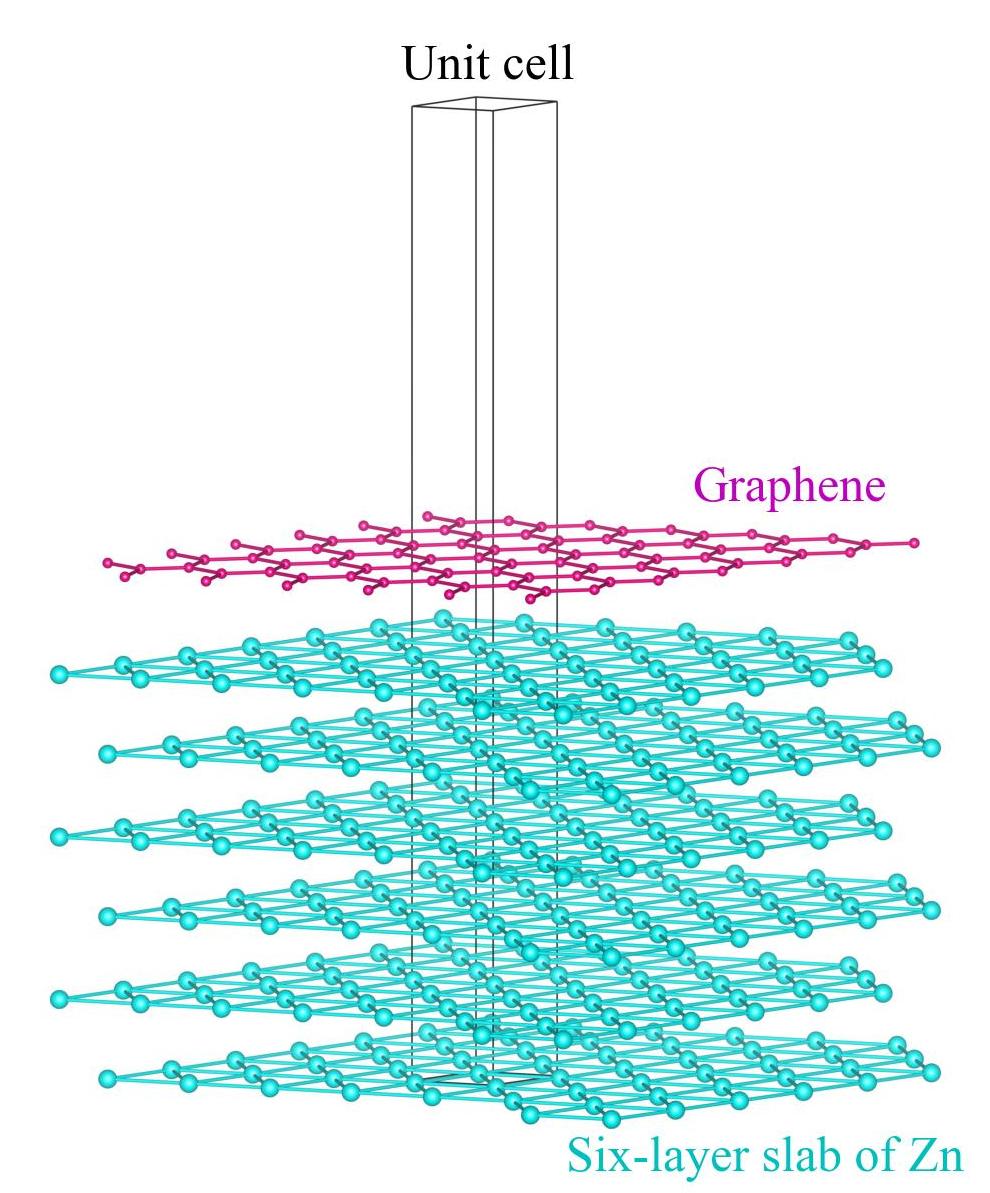}
	\caption{The atomic structure of the graphene-Zn heterojunction for first-principles calculations.
	}\label{fig:1}
\end{figure}

We perform the first-principles calculations by using Quantum Espresso (QE) \cite{16}\cite{17}with norm conserving pseudopotentials\cite{30} and Perdew-Burke-Ernzerhof exchange-correlation functional\cite{31}. The D3-type Van der Waals correction is included to improve the description of C-Zn interfacial coupling\cite{32}. The plane-wave energy cutoff is set to 80 Ry. The electronic convergence criterion is 10$^{-10}$ Ry. EPC is first calculated by the EPW code on a $24\times 24\times 1$ ($6\times 6\times 1$) k(q) mesh, and then interpolated onto a $180\times 180\times 1$ ($90\times 90\times 1$) k(q) mesh. The anisotropic Eliashberg equation is solved by setting the k and q meshes both to be $60\times 60\times 1$. 

\begin{figure*}
	\includegraphics[width=2.0\columnwidth]{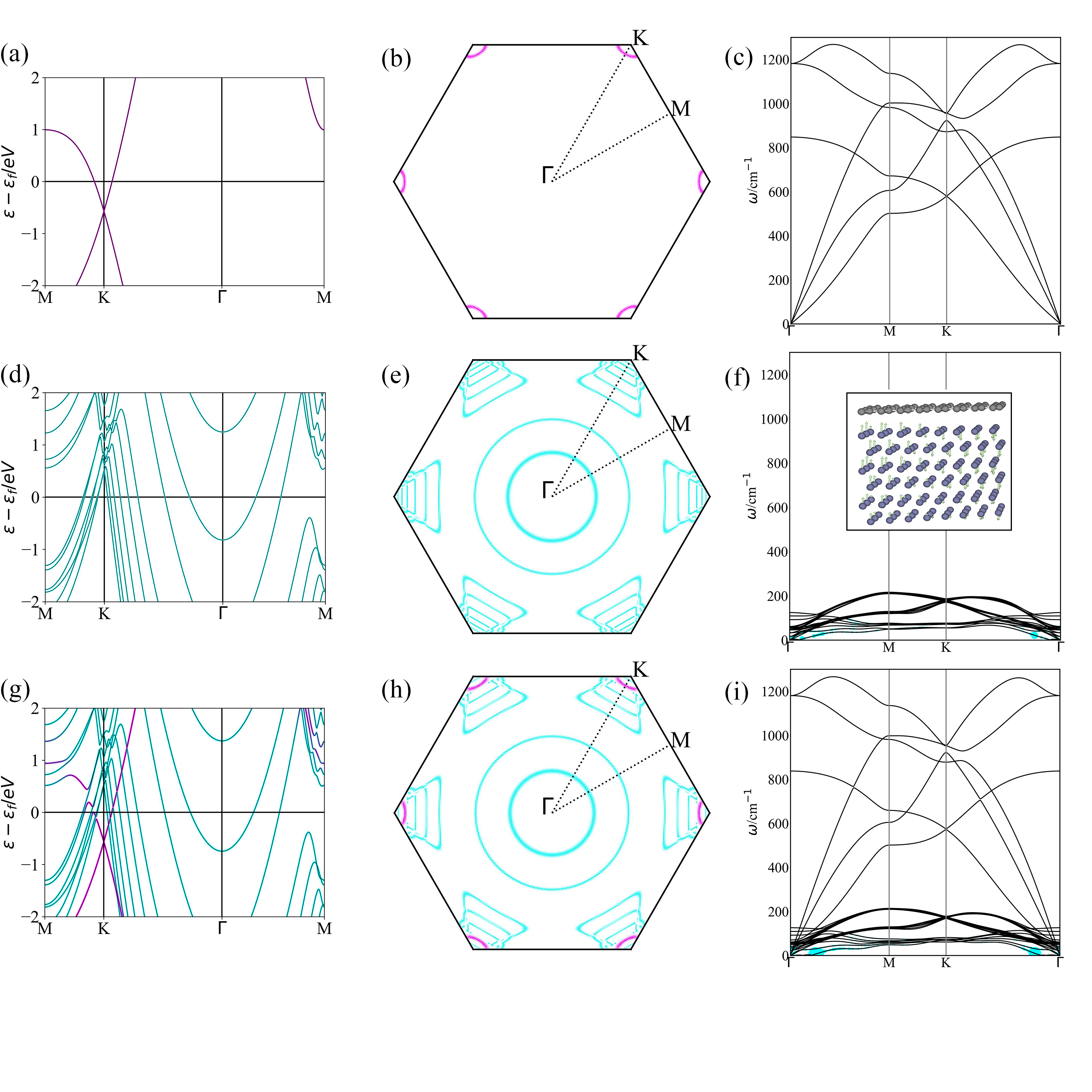}
	\caption{Electronic band structures, Fermi surfaces and phonon spectra of (a-c) graphene, (d-f) Zn slab, and (g-i) Zn/C heterostructure. Blue dots in (f) and (i) denote the EPC constant asscociated with a phonon mode. The inset of (f) visualizes the vibration mode contributing most to EPC constant in the Zn slab, as well as the Zn/C heterostructure.}\label{fig:2}
\end{figure*}

\subsection{Results}

A quick exposure of $H^{PE}$ defined in Eq. (\ref{eq:HPE}) can be visualized by comparing several key properties before and after the junction is formed. Figure 2 plots the electronic band structures, Fermi surfaces and the phonon dispersions of graphene, Zn slab and the heterojunction. The junction properties can be well tracked back to the two consisting parts, owing to a relatively weak interfacial coupling in this case. Nevertheless, electronic tunneling effects can be observed from the small hybridization gaps whenever a graphene band and a Zn band cross.  The Zn FS can be divided into $\Gamma$-centered sheets and $K$-centered sheets. The latters are most relevant to hybridizing with the graphene Dirac bands. The Fermi level of the heterojunction is determined self-consistently during the DFT loop, which indicates  0.026 electron per unit cell transferring from Zn to graphene spontaneously. For the freestanding graphene, we manually adjust the Fermi level to the same electron filling as in the junction. In the phonon spectra, the size of the blue markers reflects the phonon-resolved dimensionless EPC constant ($\lambda_{q\nu}$).  It is found that in the heterojunction [Fig. 2(i)] the EPC is dominated by the long-wave out-of-plane acoustic vibration  of the Zn atoms, which inherits the feature of the pure Zn slab [Fig. 2(f)]. Interestingly, forming a heterojuction leads to a slight enhancement of these $\lambda_{q\nu}$.

\begin{figure*}
	\includegraphics[width=2.0\columnwidth]{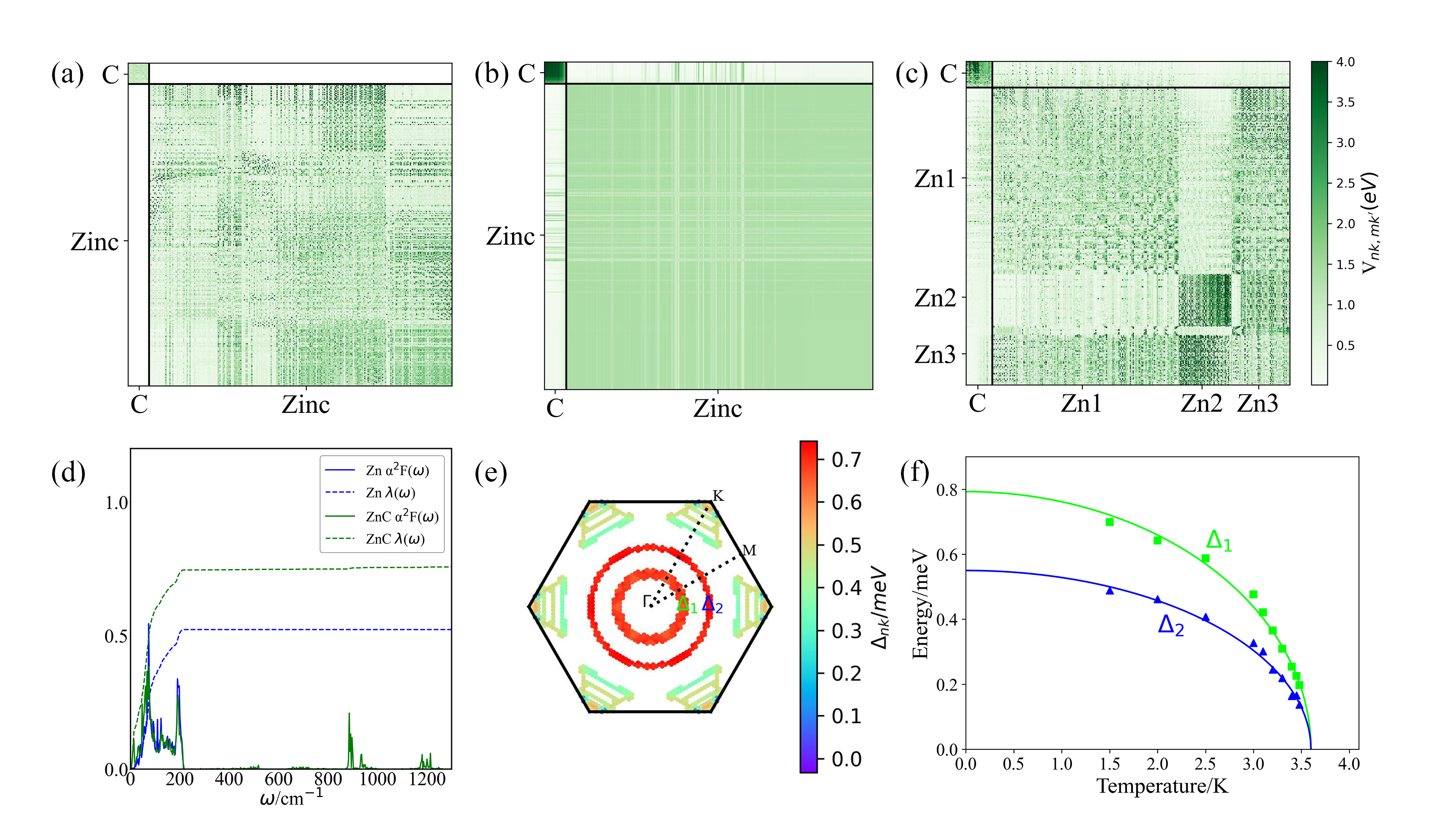}
	\caption{$V^{NS}_{n\textbf{k},m\textbf{k}'}$ matrix in (a) separated graphene and Zn,  and (b,c) the heterostructure based on the electron tunneling approximation and  the full first-principles calculation  respectively. (d) Eliashberg function $\alpha^2 F(\omega)$ with integrated EPC constant $\lambda(\omega)$, (e) SC gap on the FS at 1K. The temperature evolution of the SC gaps on the two $\Gamma$-centered FS sheets ($\Delta_1$ and $\Delta_2$) is ploted in (f) by solving anisotropic Eliashberg equations at different temperature.  
	}\label{fig:3}
\end{figure*}

Figure 3(a) plots the \textit{ab initio} $V^{NS}_{n\textbf{k},m\textbf{k}'}$ matrix when graphene and Zn are separated, and thus there is no scattering between them. We select the EPCs from scatterings between a pair of electronic states within a $\pm$ 200 meV energy window around the FS, and a 10 meV wide smearing function of the gaussian type is used to numerically replace $\delta(\epsilon_{n\textbf{k}}-E_F)$. Note that phonon-induced attraction in graphene is not weak, but since the FS density of states is low, the dimensionless EPC constant is small. 

Figure 3(b) is an estimation of the $V^{NS}_{n\textbf{k},m\textbf{k}'}$ matrix in the heterostructure by using the electron tunneling approximation [Eq. (\ref{eq:VNS1})], which introduces $V^{NS}_{N\rightarrow S}$ and $V^{NS}_{S\rightarrow N}$, giving rise to the structure of a $2\times2$ block matrix as expected in Eq. (\ref{VNS}). 

Figure 3(c) plots the \textit{ab initio} $V^{NS}_{n\textbf{k},m\textbf{k}'}$ matrix .  These FS electronic states are sorted in descending order of the projection weight in graphene, i.e. the upper left corner corresponds to the graphene dominated block. The definition of the block boundary [black solid lines in Fig. 3(c)] is chosen to be $w^S_{n\mathbf{k}}=w^N_{n\mathbf{k}}$ .  The full \textit{ab initio} $V^{NS}_{n\textbf{k},m\textbf{k}'}$ matrix displays a richer structure within the Zn block, indicating that the interface modulates the attraction potential on different sheets of the Zn FS. This type of PE is clearly beyond the scope of electron tunneling approximation.

We reduce the $V^{NS}_{n\textbf{k},m\textbf{k}'}$ matrix derived from the electron tunneling approximation [Fig. 4(b)] to a $2\times2$ dimensionless EPC matrix according to Eq. (\ref{eq:Lambda}):
\begin{eqnarray}\label{eq:level1}
\Lambda=
\left(
\begin{array}{cc}
   0.126 &  0.025  \\ 
   0.112  &  0.525
\end{array}
\right),
\end{eqnarray}
The largest eigenvalue $\lambda_{max}$=0.532, as dominated by the EPC of the Zn part. The eigenvector associated with $\lambda_{max}$ is $(0.266, 0.964)$.

Reducing the first-principles $V^{NS}_{n\textbf{k},m\textbf{k}'}$ matrix [Fig. 3(c)]  to a $2\times2$ dimensionless EPC matrix gives:
\begin{eqnarray}\label{eq:level2}
\Lambda=
\left(
\begin{array}{cc}
 0.171 &  0.031  \\ 
0.368  &  0.785 
\end{array}
\right),
\end{eqnarray}
The largest eigenvalue $\lambda_{max}$=0.803, and the associated eigenvector is $(0.504, 0.864)$. 

We can also preserve the additional structure within the Zn block, partitioning the $V^{NS}_{n\textbf{k},m\textbf{k}'}$ matrix into a $4\times4$ dimensionless EPC matrix. The block average leads to:
\begin{eqnarray}\label{eq:level3}
\Lambda=
\left(
\begin{array}{cccc}
    0.171  &  0.033   & 0.027   & 0.028 \\
    0.231   & 0.395   & 0.356   & 0.534 \\
    0.073   & 0.125   & 0.438  &  0.290 \\
    0.075    &0.159 &   0.365  &  0.137
\end{array}
\right),
\end{eqnarray}
The largest eigenvalue $\lambda_{max}$=0.900, with the eigenvector $(0.243,0.364,0.720,0.538)$. 


Figure 3(d) shows the SC gap on the FS at 1K, derived from the first-principles anisotropic Eliashberg equations. The gap size varies dramatically on different sheets of the Zn FS, in consistent with the structure of the first-principles $V^{NS}_{n\textbf{k},m\textbf{k}'}$ matrix. The proximity induced gap (0.1$\sim$0.3 meV) can be found at the graphene FS.  Figure 3(e) plots the distrubtion of the gap size. Tracing the temperature evolution of the two marked peaks ($\Delta_1$ and $\Delta_2$) determines a $T_{c}=3.6K$[Fig. 3(f)]. 


\subsection{Discussion}
For the solution of the Eliashberg equations we chose $\mu^*=0.115$. An exact determination of the $\mu^*$ value is beyond the scope of the present work. In the discussions below, we will always use this fixed $\mu^*$ value without further fine tuning, and the Eliashberg results can be regarded as a benchmark of the performance of the other approximated treatments.

By feeding $\lambda_{max}$ and $\mu^*$ into Eq. (\ref{eq:MAD}), the estimated $T_c$'s based on $\lambda_{max}$'s from Eqs. (\ref{eq:level1}), (\ref{eq:level2}), and (\ref{eq:level3}) are 0.9 K, 3.2 K and 4.0 K, respectively. The last two numbers, from the $2\times2$ or $4\times4$ partitioning of the first-principles $V^{NS}_{n\textbf{k},m\textbf{k}'}$ matrix respectively, are in reasonable agreement with the Eliashberg result $T_{c}=3.6K$. The first one, from the electron tunneling  approximation is significantly lower, but very close to the intrinsic $T_c$ of bulk Zn (0.79K) as determined in experiment~\cite{27}.  This result is understandable: in the electron tunneling approximation, the SC is essentially inherited from the isolated Zn-slab, while all the other treatments include extra interfacial effects in addion to electron tunneling.

The enhancement of SC in the heterojunction as predicted by both the first-principles $V^{NS}_{n\textbf{k},m\textbf{k}'}$ matrix and the Eliashberg results is attributed to some long-wave out-of-plane acoustic phonons[cf. Figs.2(f,i) and 3(d)]. This result is interesting, but should be interpreted with caution, because such type of vibration is sensitive to the interfacial environment. Just like the gap variation predicted by the Eliashberg equations can be easily washed out by defects and disorders in a real sample, the interfacial-enhanced SC might only occur in the ideally clean limit. Nevertheless, this result vividly demonstrates that it is possible to tune SC by controlling the interfacial structure. 

According to the eigenvector associated with $\lambda_{max}$, the electron tunneling approximation estimates the proximity induced SC gap in graphene to be $0.267/0.964\approx28\%$ of the intrinsic Zn SC gap. The estimation of the first-principles $V^{NS}_{n\textbf{k},m\textbf{k}'}$ matrix in a $2\times2$ partitioning is $0.504/0.864\approx58\%$. For the   $4\times4$ partitioning, the ratio varies between $34\%$ and $67\%$, depending on which Zn block is used as the reference. At the semi-quantiative level, these estimations predict the order of magnitude in consistent with the Eliashberg results [c.f. Fig. 3(e)].


\section{Conclusion}

In summary, we show that the power of first-principles calculation can be extended to quantify proximity-induced superconductivity. The electron tunneling approximation can be employed to significantly reduce the computational cost, putting forth a quick and convenient way to semi-quantitatively estimate the proximity induced SC gap. A full EPC calculation on the heterostructure captures further interfacial effects, such as phonon renormalization and interfacial phonon scattering, providing useful information for interfacial SC engineering.  By properly block averaging the EPC matrix as for a multi-band superconductor, a simple block average and eigenvalue analysis is found to give quantitative predictions comparable to the much more time-consuming Eliashberg equations. This methodology is expected to find general applications in the studies of interfacial SCs.

\section{Acknowledgement}
Y.L. and Z.Z. contributed equally to this work. We acknowledge support from the National Natural Science Foundation of China (11874079 ), Tsinghua University Initiative Scientific Research Program, and the open research fund program of the State key laboratory of low dimensional quantum physics (KF202103).
 
\bibliography{references}

\end{document}